# Communication Through Jamming over a Slotted ALOHA Channel

Sandeep Bhadra, Shreeshankar Bodas, Sanjay Shakkottai and Sriram Vishwanath

Wireless Networking and Communications Group

Department of Electrical and Computer Engineering

The University of Texas at Austin

{bhadra,bodas,shakkott,sriram}@ece.utexas.edu

*Short Paper/Correspondence Submission*

### Abstract

This work derives bounds on the jamming capacity of a slotted ALOHA system. A system with $n$ legitimate users, each with a Bernoulli arrival process is considered. Packets are temporarily stored at the corresponding user queues, and a slotted ALOHA strategy is used for packet transmissions over the shared channel. The scenario considered is that of a pair of *illegitimate* users that jam legitimate transmissions in order to communicate over the slotted ALOHA channel. Jamming leads to binary signaling between the illegitimate users, with packet collisions due to legitimate users treated as (multiplicative) noise in this channel. Further, the queueing dynamics at the legitimate users stochastically couples the jamming strategy used by the illegitimate users and the channel evolution.

By considering various i.i.d. jamming strategies, achievable jamming rates over the slotted ALOHA channel are derived. Further, an upper bound on the jamming capacity over the class of *all ergodic jamming policies* is derived. These bounds are shown to be tight in the limit where the offered system load approaches unity.

This research has been supported by NSF Grants ACI-0305644, CNS-0325788, CCF-0448181, CNS-0347400, CNS-0615061 and CNS-0626903. A shorter version of this paper appeared in the proceedings of The Allerton Conference on Communication, Control, and Computing, 2004.

# I. INTRODUCTION

A typical slotted ALOHA system [1], [3], [13] comprises of a collection of legitimate users following a pre-arranged strategy to gain access to resources and communicate with each other. Our work focuses on using jamming as an unconventional communication mechanism to achieve a non-zero throughput in a slotted ALOHA system. In this mechanism, an illegitimate jamming transmitter that has gained entry into a slotted ALOHA system jams legitimate transmissions, and the resulting "collisions" in the system are then detected by an illegitimate jamming receiver. Such a jamming-based communication strategy is parasitic in nature and can remain undetected without proactive effort by the legitimate entities in the slotted ALOHA system. In this work, we employ an information theoretic approach to determine upper and lower bounds on the capacity of this jamming-based communication system, under the constraint that jamming *does not* result in instability of the legitimate user queues. It is intuitively clear that the with such a constraint, the capacity of the jamming channel will converge to zero as the offered load (due to legitimate users) approaches unity. Our bounds verify this intuition, and we show that both the upper and lower bounds converge to zero as the offered load approaches unity.

A vast body of literature exists that studies the effect of illegitimate communication strategies that exploit inherent weaknesses in conventional systems. Covert communication is one such area of research where the goal of the illegitimate communication system is to exploit these weaknesses while remaining undetected by the legitimate system. A covert channel is loosely defined as an unintended or unauthorized communication path through a medium that violates the security policy of that medium. Along the lines of our jamming-based communication system, such channels are parasitic in nature, and reduce the capacity of the legitimate host channel by interfering with its communication. More formally, in a top-level characterization of covert channels, Kemmerer [11] states that necessary conditions for the existence of a covert channel are: the presence of a global resource to which both the sender and the receiver have access, a means of modifying that resource, and a method of synchronization between the receiver and the sender.

The topic of covert channels has received considerable attention among researchers in secure system design and secure source code design [14], [7], [6]. Existing results on covert channels can be divided into two major categories, *storage channels* [18] and *timing channels*. Moskowitz and Kang [14] define a storage channel as a covert channel where the covert symbol alphabet consists of asynchronous responses of a global resource (ACK/NACK responses from a processor, success/failure of a packet transmission). Shieh [17] models covert channels as finite state graphs to estimate the bandwidth (bit/s) of a covert storage channel. A covert timing channel encodes by modulating the time intervals between successive responses [14], [7], [10]. The capacity of timing channels was investigated by Anantharam and Verdú [2]. Subsequently, the capacity of covert timing channels was investigated by Giles and Hajek [7], where the authors consider the time interval information between successive transmissions of packets from a queue as a timing channel. They model this channel as an information-theoretic game between an illegitimate user who attempts to modulate these inter-arrival times and a 'jammer' who introduces random delays in the transmitted packets to arrive at bounds on $\max-\min$ and $\min-\max$ rates of mutual information in covert timing channels.

In the context of an ALOHA channel, the authors in [6] consider jamming based communication over a slotted ALOHA channel, where an FCFS based splitting algorithm is used for contention resolution [19]. They consider a scenario with a large number of users (with the aggregate arrival rate being Poisson with rate $\mu$ packets per slot), and develop two protocols for jamming based covert communication. In the procedures developed in [6], the illegitimate transmitter communicates by means of influencing the number of collisions that occur within the contention resolution period, and the illegitimate receiver uses a maximum likelihood decoder to determine the number of collisions caused by the illegitimate transmitter. They demonstrate through numerical methods that the ALOHA system can support persistent interference by the illegitimate user (using the procedures developed in [6]) without causing user packet backlogs to drift to infinity, only if the multi-access channel is lightly loaded ($\mu \approx 0.1$).

*A. Main Contributions*

In this paper, our focus is on the fundamental capacity limits of the covert ALOHA channel over the class of all ergodic jamming strategies.

- We study the information-theoretic capacity of the illegitimate system where $n$ legitimate users (where $n$ is any finite number) communicate over a slotted ALOHA channel, and for any fixed offered load $\alpha \in (0, 1)$, subject to a stability constraint on the legitimate user queues. We first derive achievable jamming rates over the slotted ALOHA channel by considering various i.i.d. jamming strategies, and where the illegitimate user has varying degrees of side-information on the channel state.
- We derive an upper bound on the jamming capacity of this channel over the class of all ergodic undetectable (*to be defined*) strategies, subject to stability constraint on the legitimate user queues. The dynamics of this system are complex because the jamming strategy of the illegitimate user influences the queueing dynamics of all the legitimate users, thus coupling the source (illegitimate user) and the channel state (the queue lengths of all the users). We also show that this upper bound is tight as the offered load approaches unity.

To obtain an upper bound, we first decouple the state of the illegitimate channel from the jamming strategy by considering a *virtual parallel channel* (which is stochastically coupled with the true channel) along with a pair of *virtual* illegitimate users. However, our construction is such that the dynamics of the virtual illegitimate users do not modify the dynamics of the virtual channel. Using our construction, we prove that the capacity of this virtual illegitimate channel is always greater than that of the true illegitimate channel and then bound it as a weighted sum of the capacities of a codeword-weight constrained Z-channel and a rate 1 error free channel.

Further details on our communication system model are given in the next section. In Section III, we present the achievable rates for jamming-based communication for a two-user system. In Section IV, we develop an upper bound on capacity in the context of a two-user system, and provide numerical results. We generalize the results to the $n$ user case in Section V.

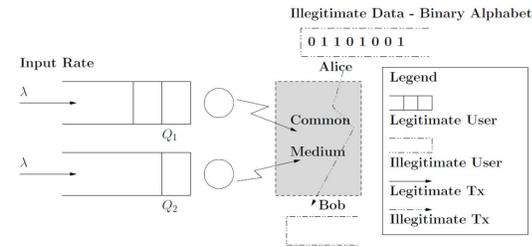
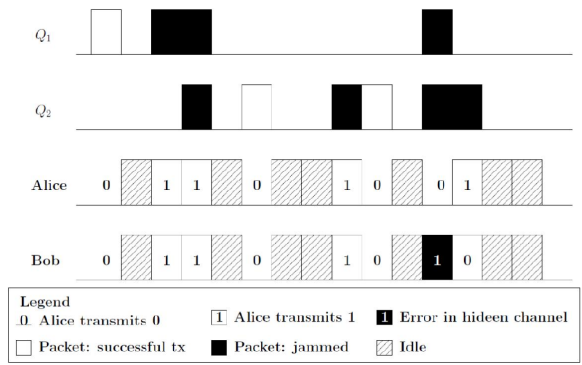

Fig. 1. System Model

Fig. 2. The Illegitimate Channel

## II. SYSTEM MODEL

In our model, we first consider the case where two legitimate users and two illegitimate users (Alice and Bob) share the common medium using slotted ALOHA[1]. Alice wishes to transmit to Bob without being detected by the system. Each legitimate user in this slotted ALOHA system is associated with a queue, with independent and identically distributed (i.i.d.) Bernoulli packet arrivals to each queue at rate $\lambda$.

A slotted ALOHA system with two legitimate users $Q_1$ and $Q_2$ is shown in Figure 1. With a slight abuse of notation, we will use $Q_i, i = 1, 2$ to denote both the users and the corresponding length of their queues. When the queue $Q_i$ is non-empty, User $Q_i$ attempts to transmit in a time-slot with probability $p$. A time-slot $j$ is said to be *active* if at least one of the users transmits a packet on the channel.

Collisions naturally occur in this system when both users $Q_1$ and $Q_2$ attempt transmission. In a regular slotted ALOHA system, such a collision is detected, and the colliding packet is then retransmitted.

We assume that the legitimate users *do not* care for the packet collisions so long as their buffers do not overflow (strictly speaking, as long as their queues do not become unstable). The legitimate users do not know how many legitimate users use the system, therefore as long as their buffers don't overflow, they consider the collisions natural. They are guaranteed by a system administrator that their buffers will not overflow under the offered load. If a legitimate user's queue becomes unstable, then (s)he complains to the system administrator, who then starts a search for potential illegitimate users. Alice and Bob have to exploit this feature to remain undetected: that is, Alice should not jam packets indiscriminately which would make the system unstable. We show that in this scenario, there is a nonzero capacity for the illegitimate channel.

Alice exploits this aspect of the system to communicate while remaining undetected, choosing signals from a binary alphabet {'0','1'}. For every '1' that Alice wishes to transmit, she causes a collision by jamming a transmission in the corresponding time-slot. Throughout this paper, we will distinguish between the terms *collision* and *jamming* according to the following convention - by *collision*, we will mean that an attempted packet transmission by either user $Q_1$ or user $Q_2$ is not successfully received; whereas, a time-slot that is *active* is said to be *jammed* if Alice transmits a '1' in that time-slot.

---

[1]We consider the generalization to the $n$ user case in Section V.

In order for the illegitimate users to remain undetected, Alice should be able to (causally) detect the presence of a legitimate packet on the channel. This can be achieved by carrier sensing or power-level detection. In practice, there exist protocols, e.g., Carrier Sense Multiple Access with Collision Avoidance (CSMA-CA) that require the transmitter to sense the carrier signal from other transmitter(s) before starting their own transmission. Also, there are commercially available cognitive radio products that sense if a particular channel is busy or not before and during its use. We idealize this situation and assume that the illegitimate transmitter can, *with probability* 1, detect the presence/absence of a legitimate transmission.

The illegitimate receiver (Bob) interprets each unsuccessful packet transmission as a '1' transmitted by Alice, and each successful transmission by the legitimate users in the system as a '0'. Neither Bob, nor the system can distinguish between collisions amongst the legitimate users and transmissions that are jammed by Alice. This indistinguishability is essential for Alice's communication to remain hidden. If Bob were granted the ability to learn to distinguish between jamming and collision, so could the legitimate system, thus exposing the illegitimate user.

Further, *Alice's jamming strategy must not make the overall system unstable* [2]. In other words, Alice's jamming strategy should be such that the queue lengths of the legitimate users should not go to infinity (a more formal description is provided in (3)). Alice's jamming policy is illustrated in Figure 2. The shaded time-slots in Figure 2 correspond to idle states when there is no activity by the legitimate users of the channel, while the solid black time-slots represent collisions in the system.

Let
$$M_i = I\{\text{channel is active in time-slot i}\}, \tag{1}$$
where $I$ is the indicator function. Thus, $M_i = 1$ if at least one of $Q_1$ or $Q_2$ transmits a packet over the common channel, and $M_i = 0$ otherwise. For each $T \in \mathbb{Z}^+$, we define the *active set*
$$A_T(\omega) = \{i : 1 \leq i \leq T, M_i = 1\} \tag{2}$$
to be the random set of active time-slots; the $\omega$ in the definition indicates that this is a random set that depends upon the queue states and the attempt probabilities at each of the legitimate user queues. However, for ease of notation, we shall drop the $\omega$ in subsequent references to this random set.

The active time-slots are indexed by the function $t(i) = \inf\{k \geq 1 : |A_k| = i\}$ which denotes the time-slot when the channel is active for the $i$-th time. The *illegitimate channel* is defined as the jamming channel between Alice and Bob. Note however, that the codewords used by Alice over this jamming channel are only transmitted (and received by Bob) over consecutive $t(i)$'s.

For the purpose of rigor, assume that whenever the channel is idle, Alice transmits a $\phi$. Thus, Alice's codewords are strings from the alphabet $\{0, 1, \phi\}$. Next, we will define $\mathcal{S}_\infty$ as the set of codeword strings of infinite length that Alice can use to jam over the illegitimate channel so that the queues $Q_1, Q_2$ are stable and ergodic. Formally, $\mathbf{x}^\infty \in S_\infty$ are such that for each $(k,l) \in \mathbb{Z}^2$, and each sample path $\omega$, the limit

$$\lim_{T\to\infty} \frac{1}{T} \sum_{i=1}^{T} I\{(Q_1(i), Q_2(i)) = (k,l)\}(\omega) \tag{3}$$

converges to a well-defined probability measure over $\mathbb{Z}^2$, where $Q_1(i), Q_2(i)$ denote the queue-lengths at time $i$.

We then define the projection (truncation) operator $\mathbb{P}_m$ operating over all strings $\mathbf{x}^n$ of length $n \geq m$ such that $\mathbb{P}_m(\mathbf{x}^n)$ is a string of length $m$ satisfying

$$(\mathbf{x}^m)_i \triangleq (\mathbb{P}_m(\mathbf{x}^n))_i, \quad \forall 1 \leq i \leq m.$$

where $(\mathbf{a})_i$ is defined as the $i$-th element in vector $\mathbf{a}$.

Formally, let $S_T$ be a set of $T$ length strings derived from $S_\infty$ under the projection operator $\mathbb{P}_T$ so that for all $\mathbf{x}^T \in S_T$, $\exists \mathbf{x}^\infty \in S_\infty$, such that $\mathbf{x}^T = \mathbb{P}_T(\mathbf{x}^\infty)$.

We can now define the (ergodic) information-theoretic *hidden capacity* over the active time-slots as follows:

$$C(\mathcal{S}) = \liminf_{T\to\infty} \sup_{\mathbf{x}^T \in \mathcal{S}_T} \frac{1}{T} I(\mathbf{x}^T; \mathbf{y}^T) \tag{4}$$

where the codeword vector $\mathbf{x}^T = (x_1, x_2, \ldots, x_T)$, each $x_i \in \{0, 1, \phi\}$ transmitted by Alice is received by Bob across the hidden channel as $\mathbf{y}^T$. The notion of the constraint sets is crucial to our definition of hidden capacity since Alice and Bob need to ensure that they remain hidden by coding such that the legitimate users are not infinitely backlogged.

Recall that we considered the $\phi$ alphabet to denote that Alice does not transmit anything over the timeslot corresponding to $\phi$ since the channel is idle at those timeslots. Bob realizes that the channel is idle and does not expect transmission by Alice. Hence the capacity in Equation (4) is

$$C(\mathcal{S}) = \liminf_{T\to\infty} \sup_{\mathbf{x}^T \in \mathcal{S}_T} \frac{1}{T} I(\mathbf{x}^{|A_T|}; \mathbf{y}^{|A_T|}) \tag{5}$$

where $\mathbf{x}^{|A_T|} = (x_{t(1)}, x_{t(2)}, \ldots, x_{t(|A_T|)})$ $x_{t(i)} \in \{0,1\}$ is the effective codeword vector transmitted by Alice and received by Bob as $\mathbf{y}^{|A_T|}$. We shall use this definition of capacity in the rest of this paper.

This paper derives analytic expressions that upper and lower bound the capacity of this illegitimate system. This capacity is less than one bit per transmission because the channel between Alice and Bob is not ideal. An error in Bob's interpretation occurs when there is a collision amongst the legitimate users in the system. A collision amongst legitimate users can only occur when more than one of them has a packet to transmit. Thus, conditioned on the event that multiple users have packets to transmit and that there is activity in the channel, the hidden channel between Alice and Bob behaves as a Z-channel [4], [9] (see Figure 3).

We assume that the illegitimate users know the offered load $\alpha = \frac{\lambda}{p\hat{p}}$, where $\hat{p} = 1-p$, and the Z-channel crossover probability $p_c$. The justification for this assumption is that if the illegitimate users do not know the offered load and start jamming with probability $\delta$, then no matter how small this $\delta$ is, there's the possibility that the system becomes unstable and the illegitimate users are exposed.

When only one of the two legitimate users has packets, there are no collisions in the legitimate channel, and the illegitimate channel reduces to an ideal error-free channel. When none of the legitimate users have packets, no transmission is possible.

III. ACHIEVABLE RATES FOR THE HIDDEN CHANNEL: THE TWO USER CASE

*A. Capacity*

The hidden channel is source dependent because the jamming strategy modifies the queues $Q_i, i = 1, 2$. It also has memory, and is constrained to ensure that the legitimate system remains stable. Conventional single letter characterizations for capacity (used for discrete memoryless channels) cannot be used in this context and hence a closed form expression in terms of channel parameters is difficult to obtain. The next sections investigate achievable rates for this channel under i.i.d. jamming strategies, and an upper bound is then used to motivate this i.i.d. jamming strategy.

*B. I.i.d. Jamming Strategies*

We define the following sets $S_{0,2} = \{(Q_1 = 0, Q_2 = 0)\}$, $S_{1,1} = \{(Q_1, Q_2) : Q_1 = 0, \ Q_2 > 0\} \cup \{(Q_1, Q_2) : Q_1 > 0, \ Q_2 = 0\}$ and $S_{2,0} = \{(Q_1, Q_2) : Q_1 > 0, \ Q_2 > 0\}$. In other words when $k$ of the 2 queues are backlogged, the process $(Q_1 Q_2)$ is said to be in state $S_{k,2-k}$. When the queue length process $(Q_1, Q_2) \in S_{2,0}$ and the channel is active, the illegitimate channel reduces to an equivalent Z-channel (see Figure 3), while for states $(Q_1, Q_2) \in S_{1,1}$ when the channel is active, the illegitimate channel reduces to a zero-error channel.

We first consider the system model as is and derive a lower bound on the capacity. We then provide the illegitimate users with side information so that they know the queue state process $(Q_1, Q_2)$ completely.

Let us denote the channel state in a time-slot $t$ by $S^t$. We consider coding/jamming policies described by a map $\mu : \mathcal{C} \mapsto [0, 1]$ where $\mathcal{C}$ is the set of channel states. Alice, then jams (i.e. transmits a '1') a transmission in an active time-slot $t(k)$ when the channel is in state $S^{t(k)} \in \mathcal{C}$ with probability $\mu(S^{t(k)})$ independent of all other events. In other-words, given the channel state, Alice uses a codebook that has been generated in an i.i.d. manner. Consequently, the expression for capacity achievable over such i.i.d. strategies follows from Equation (5) as

$$C(\mathcal{S}) = \liminf_{T \to \infty} \sup_{\mathbf{x}^{|A_T|} \in \mathcal{S}_T} \frac{1}{T} \sum_{k=1}^{|A_T|} I(x^{t(k)}; y^{t(k)}). \tag{6}$$

Observe that since the arrival rates at the queues are Bernoulli, the transmission attempt probabilities of both users are i.i.d., and Alice's coding strategy depends only on the current queue state independent of all other events, the queue length process $(Q_1, Q_2)$ is a Discrete Time Markov Chain (DTMC). Consequently, the hidden channel can be defined as a time varying channel where the channel states $\{S_{i,2-i}\}, i \in \{0, 1, 2\}$ follow a hidden Markov process. The complete transition matrix of this DTMC can be derived to show that the DTMC is aperiodic and positive recurrent for $\lambda < p\hat{p}$.

Mutual information rates of finite state Markov channels have been studied in [15], [8] for the i.i.d. coding case. A formula for mutual information for any regenerative stochastic process (including, in particular, for hidden

Markov inputs over a countable-state space Markov channel) is provided in [16]. However, the formula in [16] can only be numerically computed. In the following subsections, we derive closed-form expressions for each of the cases discussed above.

*1) Coding strategy 1:* The illegitimate users know that the hidden channel is an arbitrarily varying time-varying channel which is composed of a Z-channel (with known crossover probability $p_c$) and an error-free channel. Also, note that to retain the stability of the legitimate user queues and hence remain undetected, Alice cannot jam packets indiscriminately, but has to ensure that no more than a certain fraction $\beta$ of the packet transmissions are jammed. Since Alice does not have channel state information, she employs the policy $\mu(S^{t(k)}) = q$, for all active time-slots $t(k)$. In other words, Alice uses a state-independent i.i.d. jamming policy with jamming probability $q$.

Since the queue length process $(Q_1, Q_2)$ is a Discrete Time Markov Chain (DTMC), we can solve the global balance equations and sum over the probabilities of the relevant states to arrive at the following steady state invariant probabilities for the illegitimate channel,

$$\begin{aligned} P(S_{0,2}) = \pi_{0,2} &= \left(\frac{p\hat{q}-\lambda}{p\hat{q}}\right) \frac{p\hat{p}\hat{q}-\lambda}{p\hat{p}\hat{q}-\lambda+\lambda\hat{p}} \\ P(S_{1,1}) = \pi_{1,1} &= 2\left(1 - \frac{\lambda}{p\hat{p}\hat{q}}\right) \frac{\lambda\hat{p}}{p\hat{p}\hat{q}-\lambda+\lambda\hat{p}} \\ P(S_{2,0}) = \pi_{2,0} &= 1 - \pi_{0,2} - \pi_{1,1}, \end{aligned} \quad (7)$$

where $\hat{q} = 1 - q$. Further, with this i.i.d. jamming strategy, the stability constraint leads to the inequality $q \leq \beta$ (recall $\beta$ is an upper bound on the fraction of transmissions that can be jammed). We can now calculate $\beta$ from the global balance equations in terms of the offered load $\alpha = \lambda/p\hat{p}$ of the queues as follows,

$$\beta = 1 - \alpha.$$

to ensure that the $0 < \pi_{i,2-i} < 1$ for $i \in 0, 1, 2$ in Equation (7).

Hence the state-independent i.i.d. coding strategy for Case 1 is to find the optimal value of $q$. To obtain an expression for the capacity of this arbitrarily varying channel, we will first decompose the channel into two states $S_{1,1}$ and $S_{2,0}$ and calculate the channel capacities for a channel fixed at each of these states. Note that we exclude the state $S_{0,2}$ since there are no active time-slots in when the channel is in this state.

We define the channel-state dependent active time-slots $M_k^{(i,2-i)} = I$(at least one of the users transmits in time-slot $k|S^k = S_{i,2-i})$. Analogously, we define $A_T^{(i,2-i)} = \{k : M_k^{(i,2-i)} = 1\}$ to be the active time-slots when the channel is at state $S_{i,2-i}$.

Accordingly, define

$$C_{i,2-i}(\mathcal{S}) = \liminf_{T\to\infty} \sup_{\mathbf{x}^{|A_T|}\in\mathcal{S}_T} \frac{1}{T} \sum_{k=1}^{|A_T|} I(x_{t(k)}; y_{t(k)}|S^{t(k)} = S_{i,2-i})$$

to be the i.i.d. coding capacity of the channel fixed at state $S_{i,2-i}$. Here the constraint set $\mathcal{S}_T = \{\mathbf{x}^{|A_T|} : m(\mathbf{x}^{|A_T|}) \leq \beta A_T\}$, where $m(\mathbf{x}^{|A_T|})$ is the number of '1' symbols[2] in the vector $\mathbf{x}^{|A_T|}$.

---
[2]We henceforth denote the number of '1' symbols in a codeword as the Hamming weight of the codeword.

The illegitimate channel, given channel activity, is a zero-error channel at state $S_{1,1}$. Observe that

$$P(M_i^{(1,1)} = 1) = p.$$

Hence, from the strong law of large numbers,

$$\lim_{T \to \infty} \frac{|A_T|^{(1,1)}}{T} = p.$$

Thus $C_{1,1} = 1.p = p$.

In order to determine $C_{2,0}$, we first derive the expression for the capacity $C_z(\beta, p_c)$ for a Z-channel with binary codewords[3] constrained such that the number of '1' symbols be less than or equal to $N\beta$, and crossover probability $p_c$. From [9], the rate $R_z(u, p_c)$ of the Z-channel with cross-over probability $p_c$ for i.i.d. codes of Hamming weight $Nu$ is given by,

$$R_z(u, p_c) = H(u\hat{p}_c) - uH(\hat{p}_c) \tag{8}$$

which is maximized at

$$u_{max} = \frac{p_c^{p_c/\hat{p}_c}}{1 + \hat{p}_c p_c^{p_c/\hat{p}_c}}$$

where $\hat{p}_c = 1 - p_c$. Also, $R_z(u, p_c)$ is monotonically increasing for $u \leq u_{max}$ and monotonically decreasing for $u > u_{max}$. Thus the i.i.d. achievable capacity under the constrained Hamming weight condition for Alice is

$$C_z(\beta, p_c) = H(\gamma \hat{p}_c) - \gamma H(\hat{p}_c) \tag{9}$$

where $\gamma = \min(u_{max}, \beta)$. The optimality of i.i.d. coding for the weight constrained Z-channel follows by using similar steps as in Equations (24)–(27).

When the illegitimate channel is in state $S_{2,0}$ and the legitimate channel is active (with probability $P(M_i^{(2,0)}) = 1 - \hat{p}^2$), the corresponding channel has the capacity of the Z-channel under the weight-$\beta$ codeword constraint — thus $C_{2,0} = C_z(\beta, p_c)(1 - \hat{p}^2)$.

Then following the method outlined to derive the capacity for Arbitrarily Varying Channels from [5], we have

*Theorem 1:* The hidden-channel capacity is lower-bounded as,

$$C \geq C_z(\beta, p_c)((1 - \hat{p}^2)\pi_{2,0} + \pi_{1,1}p). \tag{10}$$

**Proof:** Since the Z-channel has lower capacity than the zero-error channel, the optimal codebook for the Z-channel can be used over a channel switching between the Z-channel and zero-error channel to achieve rate $C_z(\beta, p_c)$. Note that this codebook is transmitted only over the active time-slots which exists $((1 - \hat{p}^2)\pi_{2,0} + \pi_{1,1}p)$ fraction of the time. Hence the total rate is thinned by this fraction. ∎

---

[3]Note that we consider the Z-channel only over the active time-slots, thus we restrict the alphabet to the set {0,1}.

*2) Coding strategy 2:* The illegitimate users know that the offered load is $\alpha = \frac{\lambda}{p\hat{p}}$, where $\hat{p} = 1 - p$, and the Z-channel crossover probability is $p_c$. Since the Z-channel crossover probability is given by

$$p_c = \frac{p^2}{1-\hat{p}^2} = \frac{p}{2-p}, \qquad (11)$$

the illegitimate users can compute

$$p = \frac{2p_c}{1+p_c}. \qquad (12)$$

Also, since $\lambda = \alpha p(1-p)$, the illegitimate users can compute $\lambda$, the arrival rate for the user queues $Q_i$.

*Theorem 2:* The hidden capacity $C$ can be lower bounded by,

$$C \geq \max_{0 \leq q \leq \beta} R_z(q, \pi_{2,0} p_c)(1 - \hat{p}^2). \qquad (13)$$

**Proof:** Assume that Alice has a large interleaver present at the transmitter output and the Bob has the corresponding de-interleaver before the receiver input. Now the composite channel consisting of the interleaver, the illegitimate channel and the de-interleaver will be in state $S_{2,0}$ with probability $\pi_{2,0}$ and will have a crossover probability of $p_c$ given that the channel is in state $S_{2,0}$. Therefore, this composite channel may be considered to be a uniform Z-channel with crossover probability $p_c^1 = \pi_{2,0} p_c$ with rate $R_z(q, p_c^1)$ where $q$ is Alice's jamming probability. (Note that $\pi_{2,0}$ and therefore $p_c^1$ depends on $q$ and that we must have $q \leq \beta$ as before to ensure stability of the legitimate user queues.) The result follows by maximizing $R_z(q, p_c^1)(1-\hat{p}^2)$ over $q$. The extra factor $(1-\hat{p}^2)$ appears because the hidden channel is available only for this fraction of the total time. ∎

Remark: Coding strategy 2 is better than coding strategy 1. This is because $(1-\hat{p}^2) \geq p$, $\pi_{i,2-i} \leq 1$ and $C_z(\beta, p_c) = \max_{0 \leq q \leq \beta} R_z(q, p_c) \leq \max_{0 \leq q \leq \beta} R_z(q, \pi_{2,0} p_c)$.

*3) Achievable rate in presence of side information:* In the case where complete channel state knowledge is available to Alice, an alternate lower bound can be derived. Consider a coding scheme where Alice uses separate codebooks for each channel state. Let the probability of Alice transmitting a '1' in state $S_{2,0}$ be $q$ as before, while the probability of Alice transmitting a '1' in state $S_{1,1}$ be $w$. Finally, Alice does not transmit in the inactive queue state of $S_{0,2}$. In other words, for each active time-slot $t(k)$,

$$\mu(S^{t(k)}) = \begin{cases} q & \text{if } S^{t(k)} = S_{2,0} \\ w & \text{if } S^{t(k)} = S_{1,1} \end{cases} \qquad (14)$$

Using the same arguments as in Section III, steady state probabilities of the queues can be calculated as,

$$\pi_{0,2} = \frac{p(1-w) - \lambda}{p(1-w)} P(Q_1 = 0) \qquad (15)$$

$$\pi_{1,1} = 2\left(1 - \frac{\lambda}{p\hat{p}\hat{q}}\right)(1 - P(Q_1 = 0)) \qquad (16)$$

$$\pi_{2,0} = 1 - \pi_{0,2} - \pi_{1,1} \qquad (17)$$

where

$$P(Q_1 = 0) = \frac{(1-w)\left(-p + p^2(1-q) + pq + (1-p)p\alpha\right)}{p^2(1-q)(1-w) + (1-p)p(q-w)\alpha + p(1-q)(-1+w+(1-p)p\alpha)}$$

Then the hidden rate can be simply seen to be the sum of the rates of the Z-channel and the zero-error channel weighted by the probabilities that the illegitimate channel is in these states. The rate can then be maximized over possible values of $q$ and $w$ so as to retain the stability of the steady-state queue lengths at the legitimate users as follows:

*Theorem 3:* The achievable rate of the illegitimate channel as described in Section II, over all i.i.d. jamming policies over a legitimate channel with attempt probability $p$ and offered load $\alpha$, with complete channel state $(S_{i,2-i})$ information at the sender and receiver is given by:

$$C_2(p,\alpha) \geq \max_{0 \leq q \leq 1-\alpha,\ 0 \leq w \leq 1-\alpha+p\alpha} \pi_{2,0}(1-\hat{p}^2)R_z(q,p_c) + \pi_{1,1}pH(w). \tag{18}$$

IV. UPPER BOUND ON HIDDEN CAPACITY: THE TWO USER CASE

Upper bounds on capacity allow us to gauge the usefulness of the achievable strategies (namely i.i.d. coding) presented before. As detailed before, the channel between Alice and Bob is source dependent and has infinite memory. Thus, obtaining a good upper bound is difficult. In this section, we derive an outer bound on the hidden capacity of this system over the set of all *ergodic* jamming policies that Alice may employ. This ergodicity constraint on Alice's policy renders the problem tractable, and allows us to use relatively simple mathematical tools to arrive at upper bounds. To obtain an upper bound, we first decouple the state of the illegitimate channel from the coding strategy by considering a *virtual parallel channel*. We then prove that the capacity of this virtual illegitimate channel is always greater than that of the true illegitimate channel and then bound it as a weighted sum of the capacities of a Z-channel and a rate 1 error free channel.

*Theorem 4:* The hidden capacity $C^*$ for a slotted ALOHA system described in Section II achievable using ergodic jamming can be upper bounded as,

$$C^* \leq C_z(\bar{\beta})(1-\hat{p}^2) + p\left(\frac{1-p(1-\alpha)\alpha - \alpha^2}{1-p\alpha}\right) \tag{19}$$

where $C_z(\bar{\beta})$ is the capacity of the Z-channel with crossover probability $p_c = p^2/(1-\hat{p}^2)$ using codewords constrained to have no more than $\bar{\beta}$ fraction of 1's, with

$$\bar{\beta} = 1 - \alpha + \frac{1-p\alpha}{(1-p)\alpha^2} - \frac{(1-p)\alpha^2}{1-p\alpha}.$$

**Proof:** Consider a *virtual* channel $(Q_1^*, Q_2^*)$, defined as a stationary and ergodic process, so that $(Q_1^*, Q_2^*) = (Q_1, Q_2)$. In other words, for every legitimate packet transmitted over the true channel, there is a virtual packet transmitted over the virtual channel Let us assume that Cindy wishes to communicate with Doug secretly by jamming over this channel $(Q_1^*, Q_2^*)$, but that Cindy's transmit policy (jamming/not jamming any active time-slot) *does not* affect the dynamics of the queues. More specifically, if, in a particular time-slot, exactly one of the two legitimate users, say user 1 (in the original system) transmits a packet and Alice chooses not to jam, then whether or not Cindy jams it on the virtual channel, user 1 does not have to transmit that packet again. If Cindy chooses to induce a collision, Doug sees a collision in this time-slot and decodes it as a '1'. Similarly, if Alice jams a packet on the real channel, then whether or not Cindy jams it on the virtual channel, that packet is retransmitted by the respective user(s) on *both* the real and the virtual channels. However, the bit understood by Doug will depend upon whether there was a collision on the virtual channel. Hence, *by construction*, we couple the dynamics of the queues $Q_1^*$ and $Q_2^*$ to those of the queues $Q_1$ and $Q_2$ which are governed by mutual collisions, transmissions and jamming over the *real* channel (over which Alice and Bob communicate).

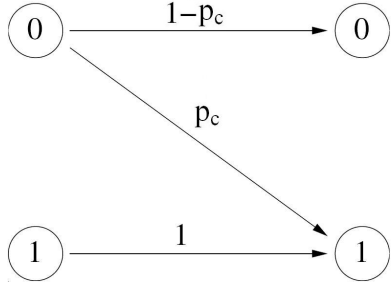

Fig. 3. The Traditional Z-Channel

Let Alice's optimal ergodic strategy be $\mathcal{A}^*$, which leads to a hidden capacity of $C^*$. From the ergodicity of $\mathcal{A}^*$ this results in steady state probabilities $\pi_{i,2-i}^*, i \in \{0,1,2\}$ corresponding to states $S_{i,2-i}, i \in \{0,1,2\}$ for $(Q_1, Q_2)$, and by our coupling construction for $(Q_1^*, Q_2^*)$ as well. Not only can Cindy replicate Alice's strategies, but since she can choose from a wider set of coding strategies (since that does not affect the dynamics of the virtual channel), the capacity that Cindy can achieve $C_c^* \geq C^*$.

Although the codewords in $\mathcal{A}^*$ might span across different states in general, the ergodicity constraint on the optimal policy implies that the fraction of time-slots jammed by Alice in each state $S_{i,2-i}$ converges to a constant $\beta_{i,2-i}^*$ defined as

$$\beta_{i,2-i}^* = \lim_{n \to \infty} \frac{1}{n} \sum_{k=1}^{n} I\{S^k = S_{i,2-i}\} I\{\text{channel is active at time-slot } k\} I\{\text{Alice transmits a '1'}\}$$

where $I\{\}$ is the indicator function and as before, $S^t$ denotes the channel state at time-slot $t$. Consequently we will apply the same codeword weight constraint $\beta_{i,2-i}^*$ to the state-dependent code that Cindy uses to communicate over the virtual channel at each state $S_{i,2-i}$.

Further, note that given queue state information, the Cindy-Doug illegitimate channel is a *discrete memoryless* time-varying channel with state side information at transmitter and receiver.

Consider a $(2^{nR}, n)$ code $\hat{\mathbf{X}}^n = \{x_i(w)\}_1^n$ over the ternary alphabet $\{0, 1, \phi\}$ transmitted over this channel with source alphabet $W$ corresponding to a state sequence (trajectory) $\mathbf{S}^n = \{S^i\}_1^n, S^i \in \{S_{k,2-k}, k \in \{0,1,2\}\}$ and received sequence $\hat{\mathbf{Y}}^n = \{y_i(w)\}_1^n$. Then following [20], we can define $C_c$ to be the capacity of the Cindy-Doug channel and $C_{i,2-i}(\beta_{i,2-i}^*)$ to be the of the Cindy-Doug channel fixed at a state $S_{i,2-i}$ under codeword constraint

$\beta^*_{i,2-i}$ as

$$C_c = \liminf_{n \to \infty} \sup_{\mathbf{X}^n \in S_n} \frac{1}{n} I(W; \mathbf{Y}^n, \mathbf{S}^n) \tag{20}$$

and

$$C_{i,2-i}(\beta^*_{i,2-i}) = \liminf_{n \to \infty} \sup_{\mathbf{X}^n : m(\mathbf{X}^n) \leq n\beta^*_{i,2-i}} \frac{1}{n} \sum_{k=1}^{|A_n^{(i,2-i)}|} I(x_{t(k)}; y_{t(k)} | S^{t(k)} = S_{i,2-i}) \tag{21}$$

respectively. We will now express the capacity of the Cindy-Doug channel $C_c$ in terms of the individual $C_{i,2-i}(\beta^*_{i,2-i})$ values.

Note that

$$nR \leq I(W; \mathbf{Y^n}, \mathbf{S^n}) \tag{22}$$
$$= I(W; \mathbf{Y^n}|\mathbf{S^n}) + I(W; \mathbf{S^n}) \tag{23}$$
$$\leq I(\mathbf{X^n}; \mathbf{Y^n}|\mathbf{S^n}) \tag{24}$$
$$= H(\mathbf{Y^n}|\mathbf{S^n}) - H(\mathbf{Y^n}|\mathbf{X^n}, \mathbf{S^n}) \tag{25}$$
$$\leq \sum_{i=1}^n H(y_i|S^i) - \sum_{i=1}^n H(y_i|x_i, S^i) \tag{26}$$
$$\leq \sum_{i=1}^n I(x_i; y_i|S^i). \tag{27}$$

The inequality in (24) follows from the assumption that the source and the state sequence are mutually independent, so $I(W; \mathbf{S^n}) = 0$, and the data processing inequality. We have inequality (25) as a consequence of the discrete memoryless nature of the channel and the inequality $H(\mathbf{Y^n}|\mathbf{S^n}) \leq \sum_{i=1}^n H(y_i|\mathbf{S^n}) \leq \sum_{i=1}^n H(y_i|S^i)$. Also observe that $I(\phi; \phi|S^i) = 0$. Dividing both sides of (27) by $n$ and using the ergodic strong law of large numbers and the definitions in Equation (21), we arrive the following bound for the capacity of the overall system with Cindy communicating to Doug:

$$C_c \leq \sum_{i,2-i} C_{i,2-i}(\beta^*_{i,2-i}) \pi^*_{i,2-i}. \tag{28}$$

We note that a similar expression as (24) is given as part of the converse proof of capacity for asymptotically block memoryless time varying channels by Médard and Goldsmith [12]. We note in passing that the sum rate in Equation (28) can be achieved by Cindy switching between codebooks corresponding to the capacity achieving code for each state $S_{i,2-i}$ without affecting the channel process $(Q_1, Q_2)$ and hence the inequality in Equation (28) can be replaced by the equality.

Next, we obtain outer bounds for $C_{2,0}(\beta^*_{2,0})$ and $\pi^*_{2,0}$. Recall that for each $T$, $A_T^{(2,0)}(\omega) = \{i : 1 \leq i \leq T, M_i^{(2,0)} = 1\}$. From the strong law of large numbers, we have that

$$\lim_{T \to \infty} \frac{|A_T|^{(2,0)}}{T} = 1 - \hat{p}^2.$$

Observe that our system model implies that for any $1 \leq j \leq T$, a transmitter Cindy, transmitting to receiver Doug over the illegitimate channel conditioned on the event that the legitimate channel exists in state $S_{2,0}$, can choose to jam a packet (i.e. transmit symbol '1') if and only if $j \in A_T^{(2,0)}$. Further, given that we are already in state $S_{2,0}$, the jamming set $A_T^{(2,0)}$ is independent of the jamming policy (codebook) employed by Cindy.

Now, for any $j \in A_T^{(2,0)}$, observe that the illegitimate channel (between Cindy and Doug) is a Z-channel with crossover probability $p_c$, where $p_c = \frac{p^2}{1-\hat{p}^2}$. Thus by concatenating the time-slots in $A_T^{(2,0)}$ (and ignoring $\{1 \leq j \leq T\} \setminus A_T^{(2,0)}$) and employing a Z-channel coding strategy over $A_T^{(2,0)}$, it follows that for any $\epsilon > 0$, $\exists T$ large enough such that,

$$C_{2,0}(\beta_{2,0}^*) \leq (C_z(\beta_{2,0}^*) - \epsilon) \frac{|A_T^{(2,0)}|}{T} \to C_z(\beta_{2,0}^*)(1 - \hat{p}^2)$$

where $C_z(\beta_{2,0}^*)$ is the channel capacity of a Z-channel with weight constraint $\beta_{2,0}^*$. For the Z-channel, it is well known that i.i.d. coding maximizes capacity [4], and hence the rate in state $S_{2,0}$ is upper bounded by $(1 - \hat{p}^2)C_z(\beta_{2,0}^*)$. In state $S_{1,1}$, given that there is activity in the legitimate channel, the channel behaves like an ideal channel (thus a trivial upper bound on $C_{1,1}(\beta_{1,1}^*)$ is 1), and the maximal rate in $S_{0,2}$ is zero. Thus, using (28) the upper bound on $C^*$ can be rewritten as

$$C^* \leq C_c \leq (1 - \hat{p}^2)C_z(\beta_{2,0}^*)\pi_{2,0}^* + p\pi_{1,1}^*. \tag{29}$$

Further (see (47) in Appendix), we have $\pi_{2,0}^* \geq \bar{\pi}_{2,0}$, where $\bar{\pi}_{2,0}$ is the steady-state probability that both user queues have packets *when no jamming is applied*. From straightforward computations, we have

$$\pi_{2,0}^* \geq \bar{\pi}_{2,0} = \frac{(1-p)\,\alpha^2}{1 - p\,\alpha}. \tag{30}$$

Hence,

$$\pi_{1,1}^* \leq \pi_{1,1}^* + \pi_{0,2}^* \leq 1 - \bar{\pi}_{2,0}. \tag{31}$$

Thus, we have that

$$\pi_{1,1}^* \leq 1 - \frac{(1-p)\,\alpha^2}{1 - p\,\alpha}. \tag{32}$$

The value of $\beta_{2,0}^*$ depends on the strategy $\mathcal{A}^*$ that Alice chooses, however we will upper bound it by $\beta_{2,0}^* \leq \bar{\beta}$ as follows. From our assumptions of ergodicity and stability of the legitimate user queues we have that

$$\begin{aligned} N\lambda &\leq N p \hat{p} \pi_{2,0}^* (1 - \beta_{2,0}^*) + Np\pi_{1,1}^* \\ &\leq N p \hat{p} \pi_{2,0}^* (1 - \beta_{2,0}^*) + Np(1 - \bar{\pi}_{2,0}). \end{aligned}$$

Thus, using the value of $\bar{\pi}_{2,0}$ from Equation (30), we can upper bound $\beta_{2,0}^*$ by

$$\beta_{2,0}^* \leq \bar{\beta} = 1 - \frac{\lambda}{p\hat{p}\pi_{2,0}^*} + \frac{1}{\bar{\pi}_{2,0}} - \bar{\pi}_{2,0} \tag{33}$$

The result now follows by observing that $\pi_{2,0}^* \leq 1$, Equations (31), and (29). ∎

We present numerical results for the achievable bound and compare it against the upper bound in Figures 4–9. The upper bound is loose everywhere except at values of $\alpha$ very close to 1. *Observe that the bound is asymptotically tight in the sense that as the offered load $\alpha \to 1$, both the upper bound and the achievable rate tend to 0.*

The bound also improves with smaller values of the transmission attempt probability $p$. These observations can be explained by noting that we have bounded $\pi_{2,0}^*$ by 1 in the $C_z(\bar{\beta})$ term of the upper bound. For smaller attempt probabilities, $\pi_{2,0}^*$ is closer to 1, even when the normalized load $\alpha$ to the queues is small. As $p$ increases, the queues at $Q_1$ and $Q_2$ are cleared promptly and hence the value of $\pi_{2,0}^*$ is much less than 1.

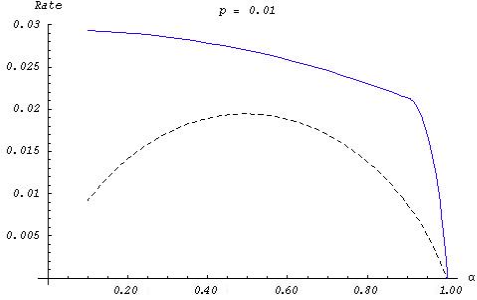

Fig. 4. Upper bound and achievable rate, p = 0.01

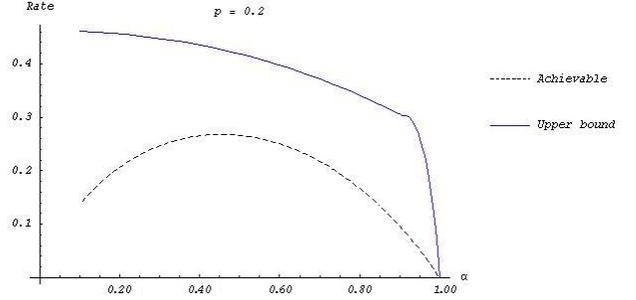

Fig. 5. Upper bound and achievable rate, p = 0.2

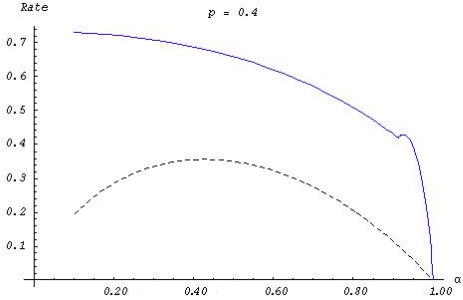

Fig. 6. Upper bound and achievable rate, p = 0.4

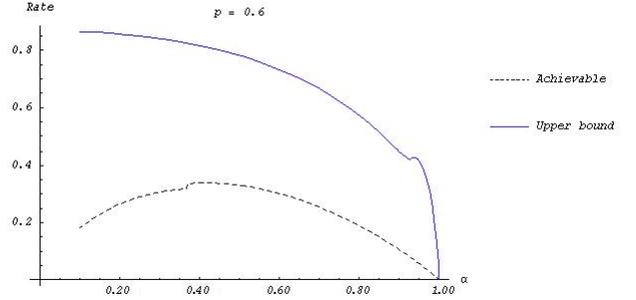

Fig. 7. Upper bound and achievable rate, p = 0.6

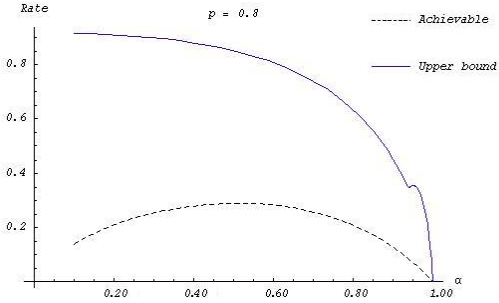

Fig. 8. Upper bound and achievable rate, p = 0.8

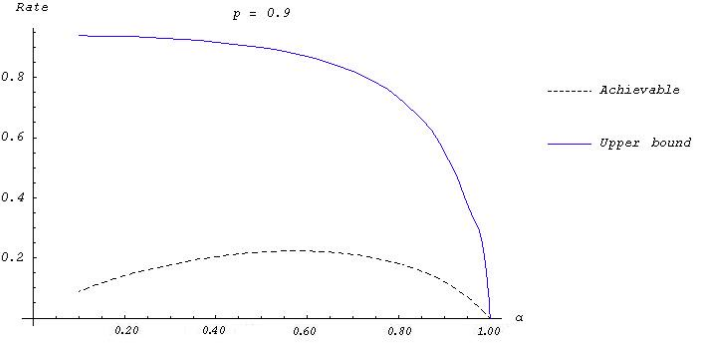

Fig. 9. Upper bound and achievable rate, p = 0.9

## V. Hidden Channels with $n$ legitimate users

Consider $n$ legitimate user queues over a common collision channel, each with homogeneous (Bernoulli) packet input rate $\lambda$. In this section we present an asymptotically (in offered load) tight upper bound to the channel capacity of the illegitimate users as a generalization of the results in Sections III and IV.

## A. Achievable Rates: The $n$ User Case

As reasoned in Section III, while the illegitimate channel depends on the state of the queue, the capacity is affected by the number of queues among the $n$ users that have packets to transmit in their buffers. For a state where $k$ of the $n$ users have packets in to transmit (non-empty buffers), we define the crossover probability of the corresponding Z-channel as

$$p_c^{(k)} = \frac{1 - (kp\hat{p}^{k-1} + \hat{p}^k)}{1 - \hat{p}^k}. \tag{34}$$

It follows that $p_c^{(k-1)} \leq p_c^{(k)} \forall k \in \{1, 2, \ldots n\}$. Correspondingly, we define $\pi_{k,n-k}$ to be the steady state probability of the channel being in any state $S_{k,n-k}$ where $k$ users out of $n$ have packets to transmit. Note that for each of the three cases of increasing illegitimate user side-information in Section III-B, the achievable rate calculation follows the same techniques as for the two user case. Due to constraints of space, we merely present the expressions for the $n$ user case with comments where necessary.

*1) Coding strategy 1:* Recall from Section III-B that the illegitimate users know the offered load $\alpha_n = \frac{\lambda}{p\hat{p}^{n-1}}$ at each legitimate user and assume that the channel is a time varying Z-channel with given crossover probability $p_c$. The hidden-channel capacity can then be bounded as

$$C \geq C_z(\beta_n, p_c) \sum_{k=1}^{n} (1 - \hat{p}^k) \pi_{k,n-k} \tag{35}$$

where $\beta_n = 1 - \alpha_n$.

*2) Coding strategy 2:* In this case, the illegitimate user views the channel as a composite Z-channel with effective crossover probability

$$\tilde{p}_c = \sum_{k=1}^{n} \pi_{k,n-k} p_c^{(k)} \tag{36}$$

resulting in an achievable rate of

$$C \geq \max_{0 \leq q \leq \beta_n} R_z(q, \tilde{p}_c)(1 - \hat{p}^n). \tag{37}$$

*3) Achievable rate in presence of side information:* We define the illegitimate user jamming probability vector $\mathbf{q} = (q_1, q_2, \ldots q_n)$ where $q_k$ is the probability that Alice jams a transmission when the system is in state $S_{k,n-k}$. Then, the achievable hidden rate under i.i.d. strategy in this case is,

$$C_n(p, \alpha) = \max_{\substack{\mathbf{q}:\forall k, \ \pi_{k,n-k} \in [0,1], \\ \sum_k \pi_{k,n-k} = 1}} \sum_{k=1}^{n} \pi_{k,n-k}(1 - \hat{p}^k) R_z(q_k, p_c^{(k)}). \tag{38}$$

## B. Upper Bound: The $n$ User Case

Analogous to the proof in Section IV, we define a weight constraint $\beta_{k,n-k}^*$ that applies on codewords that Alice (and therefore Cindy) can use for the $n$ user case. The values of $\beta_{k,n-k}^*$ depends upon the optimal strategy that Alice uses. However, we shall upper bound them as in the previous section to obtain an upper bound for the capacity.

The corresponding Z-channel capacities are denoted by $C_z^{(k)}(\beta_{k,n-k}^*)$. We trivially bound $\beta_{k,n-k} \leq 1$ for all $k < n$. For sake of uniformity of notation fix $C_z^{(0)} = 0$ and $C_z^{(1)} = H(1) = 1$. Also, following Equation (28), the capacity of the overall channel with Cindy communicating to Doug is bounded by

$$C_c^* \leq \sum_{k=0}^n \pi_{k,n-k}^* C_z^{(k)}(\beta_{k,n-k}^*)(1-\hat{p}^k)$$

For the general case of $n$ legitimate users, the Markov chain of the states of the queues of all the legitimate users is $n$-dimensional and therefore difficult to analyze. Hence we bound the values of $\pi_{k,n-k}^*$ for any transmission strategy by Alice. Consider the probabilities $\bar{\pi}_{k,n-k}$ denoting the steady state distribution of the queues *without* the presence of any illegitimate user. Using the same reasoning as (47) we have that

$$\pi_{n,0}^* \geq \bar{\pi}_{n,0} \tag{39}$$

$$\pi_{i,n-i}^* \leq \bar{\pi}_{i,n-i} \quad \forall i < n \tag{40}$$

Solving the global balance equations for $\bar{\pi}_{n,0}$, we have

$$\bar{\pi}_{n,0}^* \geq \left[1 + \frac{p\hat{p}^{n-1}}{\lambda(1-(n-1)p\hat{p}^{n-2})}\right]^{-1}. \tag{41}$$

Also, since

$$\sum_{j=0}^n \pi_{j,n-j}^* = 1 - \pi_{n,0}^* \leq 1 - \bar{\pi}_{n,0},$$

we have that,

$$\sum_{k=0}^{n-1} \pi_{k,n-k}^* C^{(k)}(1)_z \leq (1-\bar{\pi}_{n,0}) C_z^{(1)}.$$

We now bound $\beta_{n,0}^*$ in a technique similar to that used in Section IV. Observe that for stability we must have that

$$\lambda \leq \sum_{i=0}^n \pi_{n-i,i}^* p\hat{p}^{n-i-1}(1-\beta_{n-i,i}^*).$$

Trivially bounding $\beta_{n-i,i}^*$'s for $i > 0$ by 1, and using the inequalities in Equations (39), we bound

$$\beta_{n,0}^* \leq \bar{\beta}_n = 1 - \frac{\lambda}{p\hat{p}^{n-1}} + \frac{\sum_{i=1}^n \bar{\pi}_{n-i,i}(n-i)\hat{p}^{-i}}{\bar{\pi}_{n,0}} \tag{42}$$

Thus Alice's hidden capacity is bounded by,

$$\begin{aligned} C &\leq C_c \\ &\leq \sum_{k=1}^n \pi_{k,n-k} C_z^{(k)}(\beta_{k,n-k}^*)(1-\hat{p}^k) \\ &\leq \sum_{k=0}^{n-1} \pi_{k,n-k}^* C_z^{(k)} + \pi_{n,0}^* C_z^{(n)}(\beta_{n,0}^*)(1-\hat{p}^n) \\ &\leq (1-\bar{\pi}_{n,0}) + C_z^{(n)}(\bar{\beta}_n)(1-\hat{p}^n) \end{aligned} \tag{43}$$

*Theorem 5:* The hidden capacity $C^{(n)}$, for a slotted ALOHA system described in Section II with $n$ legitimate users, achievable using ergodic jamming can be upper bounded as,

$$C^{(n)} \leq \left(1 - \left[1 + \frac{p\hat{p}^{n-1}}{\lambda(1-(n-1)p\hat{p}^{n-2})}\right]^{-1}\right) + C_z^{(n)}(\bar{\beta}_n)(1-\hat{p}^n) \tag{44}$$

where $C_z^{(k)}(\bar{\beta}_n)$ is the capacity of the Z-channel for codes constrained to have less than $\bar{\beta}_n$ fraction of '1's in each codeword corresponding to a crossover probability of $p_c^{(k)}$.

**Proof:** Follows from the inequalities (41) and (43). ∎

Observe that as the offered loads approaches unity (i.e. as $\lambda \to p\hat{p}^{n-1}$), each $\bar{\pi}_{i,n-i} \to 0$ for $i < n$ in Equation (42) while $\bar{\pi}_{n,0} \to 1$. Thus $\bar{\beta}_n \to 0$ and hence $C_z^{(n)}(\bar{\beta}_n) \to 0$. Hence $C^*$ converges to 0 as the load approaches 1, and is thus asymptotically tight to the i.i.d. coding rate for the $n$ user case.

## VI. CONCLUSION

The setting studied in this paper is of two illegitimate users - a transmitter and a receiver, communicating with each other by exploiting the resources of a slotted ALOHA system. The illegitimate pair communicate by jamming legitimate transmissions while striving to remain undetected by the legitimate slotted ALOHA system. In this paper, we find that a closed-form characterization of the information-theoretic capacity of the illegitimate communication system is extremely difficult, and hence find lower and upper bounds on capacity. We employ i.i.d. coding strategies under varying side-information assumptions to determine lower bounds. Next, we employ constrained decoupling arguments to determine upper bounds, and finally, we compare the upper and lower bounds. We find that, in the limit when the offered load tends to unity (and the capacity to zero), our upper and lower bounds coincide.

## APPENDIX

Consider two sets of queue length processes $(Q_1^U, Q_2^U)$ and $(Q_1^J, Q_2^J)$, with identical arrival processes $A_k^U(n) = A_k^J(n), k = \{1,2\}$, to each queue over any fixed interval of time-slots $n = 1, 2, \ldots, N$, and with identical initial state (i.e. $Q_1^J(1) = Q_1^U(1)$ and $Q_2^J(1) = Q_2^U(1)$). The process $(Q_1^U, Q_2^U)$ corresponds to the scenario where two users compete to access a shared (slotted) channel and *no illegitimate jamming* occurs over this channel. In other words, collisions occur over this channel only due to simultaneous attempts due to the two legitimate users. On the other-hand, $(Q_1^U, Q_2^U)$ corresponds to the scenario where two users compete to access a shared (slotted) channel and *illegitimate jamming* occurs over this channel. Thus, collisions could occur over this channel either due to collisions by these legitimate users, or due to a jammer (Alice) who could employ an arbitrary jamming strategy. At each time-slot, for either scenario (with or without jamming), we assume that each of the user attempts to transmit independently with probability $p$, irrespective of whether the queue has packets or not. Note that when the queue is empty, a decision to attempt does not affect the system dynamics. However, this enables us to sample-path-wise couple the two queueing systems.

Consider any system sample path corresponding to a sequence of arrivals and transmission attempts (which are identical to both $(Q_1^U, Q_2^U)$ and $(Q_1^J, Q_2^J)$). We first show that for all $n$, we have

$$\begin{aligned} Q_1^U(n) &\leq Q_1^J(n) \\ Q_2^U(n) &\leq Q_2^J(n). \end{aligned} \qquad (45)$$

We see this by contradiction. Let $l + 1 \in \mathbb{N}, 1 \leq l \leq N$ be the first time slot where (45) fails. In other words, $Q_1^U(l) \leq Q_1^J(l)$ $Q_1^U(l) \leq Q_1^J(l)$ and $Q_2^U(l) \leq Q_2^J(l)$, but (without loss of generality, say) $Q_1^U(l+1) > Q_1^J(l+1)$.

Since arrival and transmission attempts are identical in both the jammed and the unjammed queues, if queue $Q_1^J$ transmits a packet successfully (i.e. no collision occurs) the same should be true for queue $Q_1^U$. Thus, $Q_1^U(l+1) = Q_1^U(l) + A^U(l+1) - I\{Q_1^U(l) > 0\}$ and $Q_1^J(l+1) = Q_1^J(l) + A^J(l+1) - I\{Q_1^J(l) > 0\}$. However, since $Q_1^U(l) \leq Q_1^J(l)$, $I\{Q_1^U(l) > 0\} \leq I\{Q_1^J(l) > 0\}$, we have $Q_1^U(l+1) \leq Q_1^J(l+1)$ which leads to a contradiction of our hypothesis. Thus (45) is true for all $n$.

The relation
$$\sum_{n=1}^{N} \frac{1}{N} I\{Q_1^J(n) > 0, Q_2^J(n) > 0\} \geq \sum_{n=1}^{N} \frac{1}{N} I\{Q_1^U(n) > 0, Q_2^U(n) > 0\}. \tag{46}$$
follows immediately from (45).

Considering the ergodic jamming policy $\mathcal{A}^*$ used by the illegitimate transmitter in Section IV, we can use the ergodic theorem to conclude that as $N \to \infty$, (46) converges to,
$$\pi_{2,0}^* \geq \bar{\pi}_{2,0}. \tag{47}$$